\documentclass[12pt]{article}
\usepackage{hyperref}
\usepackage{cite}
\usepackage{color}
\usepackage{graphicx}

\makeatletter
\@addtoreset{equation}{section}

\makeatletter
\renewcommand\section{\@startsection {section}{1}{\z@}%
                                   {-3.5ex \@plus -1ex \@minus -.2ex}%nn
                                   {2.3ex \@plus.2ex}%
                                   {\normalfont\large\bfseries}}
\renewcommand\subsection{\@startsection{subsection}{2}{\z@}%
                                     {-3.25ex\@plus -1ex \@minus -.2ex}%
                                     {1.5ex \@plus .2ex}%
                                     {\normalfont\bfseries}}

\def\baselinestretch{1.2}
\newcommand{\be}{\begin{equation}}
\newcommand{\ee}{\end{equation}}
\newcommand{\beq}{\begin{eqnarray}}
\newcommand{\eeq}{\end{eqnarray}}

\newcommand{\gone}[1]{{}}

\newcommand{\re}{{\rm Re}}
\newcommand{\im}{{\rm Im}}

\begin{document}
\begin{titlepage}
\begin{flushright}
MAD-TH-11-02
\end{flushright}
%\vspace{12 mm}

\vfil
%vfil

\begin{center}

{\bf \large 
Quantization of charges and fluxes in  warped Stenzel geometry
}

\vfil

Akikazu Hashimoto$^a$ and Peter Ouyang$^b$

\vfil

$^a$ 
Department of Physics, University of Wisconsin, Madison, WI
53706, USA

$^b$ 
Department of Physics, Purdue University, West Lafayette, IN 47907, USA

\vfil

\end{center}

%%%%%%%%%%%%%%%%%%%%%%%%%%%%%%%%%%%%%%%%%%%%%%%%%%%%%%%%%%%%%%%%%%%%%%%%%%%%%%%%%%%%%%%
\begin{abstract}
\noindent We examine the quantization of fluxes for the warped Stiefel
cone and Stenzel geometries and their orbifolds, and distinguish the
roles of three related notions of charge: Page, Maxwell, and brane.
The orbifolds admit discrete torsion, and we describe the associated
quantum numbers which are consistent with the geometry in its large
radius and small radius limits from both the type IIA and the M-theory
perspectives.  The discrete torsion, measured by a Page charge, is
related to the number of fractional branes.  We relate the shifts in
the Page charges under large gauge transformations to the
Hanany-Witten brane creation effect.

\end{abstract}
%%%%%%%%%%%%%%%%%%%%%%%%%%%%%%%%%%%%%%%%%%%%%%%%%%%%%%%%%%%%%%%%%%%%%%%%%%%%%%%%%%%%%%%%%
\vspace{0.5in}

\end{titlepage}
\renewcommand{\baselinestretch}{1.05}  %Line spacing
%%%%%%%%%%%%%%%%%%%%%%%%%%%%%%%%%%%%%%%%%%%%%%%%%%%%%%%%%%%%%%%%%%%%%%%%%%%%%%%%%%%%%%%%%%%%%

\section{Introduction}

Recently, a holographic duality for superconformal Chern-Simons-Matter
theories in 2+1 dimensions with ${\cal N}=6$ and ${\cal N}=8$
supersymmetry was proposed \cite{Aharony:2008ug,Aharony:2008gk}. These
field theories have $U(N)_k \times U(N+l)_{-k}$ product gauge symmetry
(where the subscripts refer to the Chern-Simons level associated with
each gauge group) and bifundamental matter fields.  In the large $N$
limit, the field theory has a dual gravity description in terms of
M-theory as $N$ M2-branes on the orbifold $C^4/Z_k$ (where the
orbifold acts by rotating each of the complex planes by an angle $2
\pi/k$ simultaneously) and $l$ fractional branes.  The supergravity
solution corresponding to this brane system is $AdS_4 \times S^7/Z_k$,
and the quantum number $l$ is encoded in the discrete torsion of the
$H_3(Z) = Z_k$ homology group of $S^7/Z_k$.

The M-theory background can also be described in type IIA supergravity
by dimensionally reducing along the Hopf-fiber of $S^7/Z_k$. In this
description, the geometry has the form $AdS_4 \times CP^3$ and is the
effective description when $1 \ll N \ll k^5$. Homologically, $CP^3$ is
very different from $S^7/Z_k$, particularly in that $CP^3$ has no
discrete torsion cycles, but it does possess integral homology. Even
for this simple example, the relationship of the spectrum of charges
and fluxes in the M-theory and the type IIA descriptions is subtle.

One way to gain some perspective on the physical meaning of the
defining data of the gravity side of these correspondences is to
realize the superconformal field theory as either the UV or IR fixed
point of a holographic renormalization group flow.  For example, a
superconformal Chern-Simons theory can arise as the IR fixed point of
an RG flow from a Chern-Simons-Yang-Mills-Matter theory
\cite{Hashimoto:2008iv,Aharony:2009fc}. Several related realizations
have also been considered \cite{Hashimoto:2010bq}.  These
renormalization group flows are dual to transverse geometries which
differ from $R^8$, and many of these constructions have the
interesting property that the dual geometry admits a normalizable
4-form.  In M-theory, this allows one to introduce a nontrivial 4-form
flux.  The freedom of tuning the 4-form flux has a specific
interpretation in terms of tuning the parameters of the dual field
theory, and in some examples, one can explore dynamical features such
as phase transitions in the low energy effective field theory from the
geometry of the supergravity dual
\cite{Aharony:2009fc,Hashimoto:2010bq}.

In this article, we investigate the duality of ${\cal N}=2$
Chern-Simons quiver theories dual to $AdS_4 \times V_{5,2}/Z_k$ where
$V_{5,2}$ is a homogeneous Sasaki-Einstein seven-manifold. This
duality was originally considered by Martelli and Sparks in
\cite{Martelli:2009ga}. On the field theory side, it generalizes the
model of ABJM by adding a chiral multiplet in the adjoint
representation to each factor of the $U(N)_k \times U(N+l)_{-k}$ gauge
group. The gravity dual description can be deformed in the IR, giving
rise to a geometry known as the warped Stenzel metric. At the moment,
little is known about the field theory interpretation of this IR
deformation. In order to facilitate in its interpretation, it is
useful to enumerate the the discrete and continuous parameters
associated with this system. This is related to the problem of
quantizing charges and fluxes in the gravity dual.

In type IIA (and IIB) supergravity, there is a well-known subtlety in
imposing charge quantization, which arises in the example studied in
this paper.  The $V_{5,2}/Z_k$ geometry, reduced to IIA along the
$U(1)$ isometry along which the $Z_k$ acts, is a space $M_2$ which has
the same homology structure as $CP^3$ \cite{Martelli:2009ga}; in
particular there is a nontrivial 4-cycle.  Now, one might want to
quantize the four-form flux through this cycle, but the natural
gauge-invariant four-form
\be \tilde F_4 = d A_3 + H_3 \wedge A_1 \ee
is not closed, and therefore its integral through the 4-cycle is not
conserved and cannot be quantized.  A similar issue arises in the flux
of $* \tilde F_4$ through $M_2$. These apparent difficulties have also
appeared in earlier examples considered in
\cite{Aharony:2009fc,Hashimoto:2010bq} and their resolution is well
understood.  The four-form flux satisfies a modified Bianchi identity,
\be d \tilde F_4 = -H_3 \wedge F_2 \ee
so to define a conserved charge we should not integrate $\tilde{F}_4$
but a modified flux which is chosen to be closed:
\be Q_4^{Page} = {1 \over (2 \pi l_s)^3 g_s } \int (-\tilde F_4 + B
\wedge F_2) \label{q4page}\ .\ee
This new charge, known as the Page charge, is one of the three subtle
notion of charges identified by Marolf \cite{Marolf:2000cb}. The three
charges being referred here are the Maxwell charge, brane charge, and
the Page charge, and they can take distinct values in gauge theories
involving Chern-Simons terms as is the case for type IIA
supergravity. Each of these charges respects some, but not all, of the
properties commonly associated with charges in simpler contexts: gauge
invariance, conservation, localization, and integer quantization. Page
charge turns out to respect conservation, localization, and integer
quantization, but fails to be invariant with respect to large gauge
transformations which shift the period of $B_2$. This ambiguity is
precisely what is required to interpret the Hanany-Witten brane
creation effects in the brane construction of these models is
intimately connected to the duality of the field theory.

In this article, we will analyze the quantization of fluxes in $AdS_4
\times V_{5,2}/Z_k$ geometry and its Stenzel deformation from the type
IIA perspective, and relate the gauge ambiguity to Hanany-Witten brane
creation effects along the lines of
\cite{Aharony:2009fc,Hashimoto:2010bq}. In \cite{Martelli:2009ga}, it
was argued that the Stenzel deformation is incompatible with the
presence of discrete torsion which gives rise to a non-vanishing value
of $l$ in $U(N)_k \times U(N+l)_{-k}$.  On the contrary, we find that
some values of $l$ are allowed, and explain the source of this
apparent discrepancy.  We will also examine the compatibility of the
IIA and the M-theory perspectives.

\section{Stiefel, Stenzel, and the ${\cal N}=2$ Chern-Simons-Quiver theory}

In this section, we briefly review the construction of ${\cal N}=2$
Chern-Simons-Quiver theory, its gravity dual, and its Stenzel
deformation. We closely follow the presentation of
\cite{Martelli:2009ga}.

\subsection{Stiefel cone}
\label{sec2.1}
The starting point is a non-compact Calabi-Yau 4-fold 
\be z_0^n + z_1^2+z_2^2+z_3^2 + z_4^2 = 0 \label{curve} \ee
where we take $n=2$. This geometry is a cone whose base is a
Sasaki-Einstein seven manifold $V_{5,2}$, also known as the Stiefel
manifold. Had one taken $n=1$, the geometry of the Calabi-Yau 4-fold
would have been $R^8$ which is formally a cone over $S^7$.  For $n
>2$, the geometry is not a cone over a Sasaki-Einstein manifold
\cite{Martelli:2009ga}.

When M2 branes are placed at the tip of the cone, we obtain a warped
geometry $AdS_4 \times V_{5,2}$.  The base $Y_2=V_{5,2}$ has a torsion
3-cycle $H_3(Y_2,Z)=Z_2$.

The $Z_k$ orbifold is taken on the $U(1)_b$ isometry which rotates
\be (z_0,z_1+z_2, z_3+z_4, z_1-z_2,z_3-z_4) \ee
with weights $(0,1,1,-1,-1)$.  On $Y_2/Z_k$, this changes the torsion
group from $Z_2$ to $H_3(Y_2/Z_k,Z)=Z_{2k}$, so
\be {1 \over (2 \pi l_p)^3} \int_{\Sigma_3} C_3 = {l -k \over 2 k} \ee
for $\Sigma_3$ which generates $H_3(Y_2/Z_k)$. Here we have shifted
$l$ by $k$ compared to what is written in (3.26) of
\cite{Martelli:2009ga}. Both $l$ and $k$ are integers so this shift is
a matter of convention in describing the supergravity background.

When reduced to IIA along the $U(1)_b$ direction parametrized by
$\gamma$, the Sasaki-Einstein space $Y_2/Z_k$ decomposes into
\be ds^2(Y_2/Z_k) = ds^2(M_2) + {w \over k^2}(d \gamma + kP)^2 \ee
and the IIA string frame metric becomes
\be ds^2 = \sqrt{w} {R^3 \over k} \left({1 \over 4} ds^2(\mbox{AdS}_4)+ds^2(M_2)^2\right) \ee
with
\be F_2 = k g_s l_s \Omega_2, \qquad \Omega_2 = d P \  . \ee

Since
\be C_3 = A_3 + B_2 \wedge d \gamma \ee
with this dimensional reduction, $B_2$ turns out to have the period
\be {1 \over (2 \pi l_s)^2} \int B_2 = {l \over 2k} - {1 \over 2} \ . \ee

\subsection{Brane construction and the Hanany-Witten effect}

The field theory dual is conjectured in \cite{Martelli:2009ga} to
arise from the low energy limit of a network of D3-branes, an NS5-brane and
a $(1,k)$ 5-brane in type IIB on $R^{1,2} \times S^1 \times R^2\times
C^2$. The D3-branes wind along $R^{1,2} \times S^1$. The NS5 is
extended along $R^{1,2}$, one of the $R$ in $R^2$ and along the curve
$w_1=- i w_0^2$ where $C^2$ is parametrized by $(w_0,w_1)$. The
$(1,k)$ 5-brane is extended along $R^{1,2}$, a line at an angle
relative to the NS5-brane in $R^2$, and along $w_1=i w_0^2$ in
$C^2$. There may also be fractional D3-branes stretching between the
NS5 and the $(1,k)$ 5-brane at $(w_0,w_1)=(0,0)$.

In a brane configuration of this type, the Hanany-Witten brane
creation effect occurs when one of the 5-branes are moved around the
circle $S^1$ keeping the other 5-brane fixed. If there were $N$
integer and $l$ fractional branes to start with, moving the 5-brane
once around the circle will give rise to a shift
\beq N & \rightarrow & N + l  \cr
l & \rightarrow & l + 2k \ . 
\eeq

\subsection{Stenzel Deformation}
\label{sec2.3}

In this subsection, we will briefly review the IR deformation of the
Stiefel cone.  As an algebraic curve, it amounts to deforming
(\ref{curve}) to
\be z_0^2 + z_1^2+z_2^2+z_3^2 + z_4^2 = \epsilon^2 \label{stenzeleq}\ . \ee
The tip of the cone is blown up by an $S^4$ parametrized by 
\be \sum_{i=0}^4 (\re z_i)^2 =\epsilon^2, \qquad \im z_i = 0 \ee
and the full geometry can be viewed as the cotangent bundle over
$S^4$. This geometry is also known as the Stenzel geometry
\cite{Stenzel} and admits an explicit metric \cite{Cvetic:2000db}. In
the notation adopted in \cite{Klebanov:2010qs}, the metric takes the
form

\be ds^2 = c^2( dr^2 +  \nu^2) + a^2 \sum_{i=1}^3 \sigma_i^2 + b^2 \sum_{i=1}^3 \tilde \sigma_i^2 \ee
where
\beq 
a^2(r) &=& 3^{-1/4} \lambda^2 \epsilon^{3/2}  (2 + \cosh 2r)^{1/4} \cosh r, \cr
b^2(r) &=& 3^{-1/4} \lambda^2 \epsilon^{3/2}  (2 +\cosh 2r)^{1/4} \sinh r \tanh r \cr
c^2(r) &=& 3^{3/4} \lambda^2 \epsilon^{3/2}(2+\cosh 2r)^{-3/4} \cosh^3 r \eeq
and $\nu$, $\sigma_i$, and $\tilde \sigma_i$ are left-invariant
one-forms of the coset $SO(5)/SO(3)$ (for which a nice explicit basis
appears in \cite{Klebanov:2010qs}.)

At $r=0$, the geometry collapses to an $S^4$. At large $r$, the
geometry asymptotes to a cone over $V_{5,2}$.  Formally, this geometry
is similar to the deformed $B_8$ space \cite{Gibbons:1989er} which
collapses to an $S^4$ near the tip, and asymptotes to cone over a
squashed 7-sphere, but there are a few important differences. One is
the fact the $Z_k$ orbifold along the $U(1)_b$ of the Stenzel geometry
has fixed points at antipodal points of $S^4$ at $r=0$. We will
comment on other differences below.

One important feature of the Stenzel geometry is that it admits a
self-dual 4-form which can be written, explicitly, as
\beq G_4 &=& m \left\{ {3  \over \epsilon^3 \coth^4{r \over 2}} \left[ a^3 c \nu \wedge \sigma_1 \wedge \sigma_2 \wedge \sigma_3 + {1 \over 2} b^3 c d r \wedge \tilde \sigma_1 \wedge \tilde \sigma_2 \wedge \tilde \sigma_3 \right] \right. \cr
&& \left. + {1 \over 2 \epsilon^3 \coth^4{r \over 2}} \left[{1 \over 2} a^2 b c \epsilon^{ijk} d r \wedge \sigma_i \wedge \sigma_j \wedge \tilde \sigma_k + a b^2 c \epsilon^{ijk} \nu \wedge \sigma_i \wedge \tilde \sigma_j \wedge \tilde \sigma_k \right]\right\} \ . 
\label{g4}
\eeq

Because the four-form is self-dual, and the background geometry is
Calabi-Yau, one can turn on this flux in eleven-dimensional
supergravity without breaking supersymmetry \cite{Becker:2001pm}.
Moreover, it gives rise to a solution where the background geometry is
unmodified except for the presence of a warp factor $H$, as in the
standard warped product ansatz
\beq ds^2 &=& H^{-2/3} (-dt^2+dx_1^2 + dx_2^2) + H^{1/3} ds_8^2 \cr
F_4 & = & dt \wedge dx_1 \wedge dx_2 \wedge d \tilde H^{-1} + m G_{4} \ . \label{ansatz} \eeq
The warp factor itself can be determined by solving the four-form field equation,
\be d * G = {1 \over 2} G \wedge G \ , \ee
where in general there can be additional source terms due to the
presence of explicit M2-branes.

\section{Quantization of fluxes in Stiefel cones and Stenzel space}

Let us now consider the quantization of fluxes in the warped Stiefel
cones and Stenzel geometries in order to identify the discrete
parameters characterizing the background. There are two guiding
principles which we follow in carrying out the quantization. One is
that quantized fluxes should be invariant under deformation of
Gaussian surfaces unless the discrete unit of charge crosses the
surface. The other is for the quantization condition to be invariant
under string dualities.

\subsection{Review of Maxwell, brane, and Page charges}

We begin by considering the quantization of fluxes for the Stenzel
geometry in the IIA description. While the IIA description of the
Stenzel geometry is singular near the core, one still expects Gauss
law considerations to lead to a consistent picture far away from the core
region, where the geometry looks essentially like the warped Stiefel
cone.

The relevant fluxes to consider then are the flux of $\tilde F_4$ through
the generator of $H_4(M_2,Z)$ and the flux of $*\tilde F_4$ through
the six cycle $M_2$. As we mentioned earlier, however, these fluxes
depend on the radius $r$ at which we identify the base $M_2$ for the
background in consideration.

The resolution to these apparent difficulties is the realization that
one is dealing with a situation where the Maxwell, brane, and Page
charge are distinct from one another, and that care is required in
applying  quantization conditions on the appropriate charge.

Let us recall the specific definition of three charges. In type IIA
supergravity, the four form $\tilde F_4 =d A_3 + H_3 \wedge A_1$ is
gauge invariant and well defined but is not closed and does not
respect Gauss' law. One can nonetheless compute the period of $\tilde
F_4$ on the generator of $H_4(M_2,Z)$ in the $r \rightarrow \infty$
limit. This defines the Maxwell charge. In contrast, the period of
Page flux $-(\tilde F_4 +B_2 \wedge F_2)$ on $H_4(M_2,Z)$ is
independent of $r$, although it is ambiguous with respect to large
gauge transformation of $B_2$. This quantity defines the Page
charge. Finally, the amount of charge carried by a brane source
through its Wess-Zumino couplings defines the brane charge. Brane
charge includes the contribution of lower-brane charges from the
pull-back of the NSNS 2-form in the Wess-Zumino coupling. Therefore,
if the background contains a non-uniform NSNS 2-form $B_2$, the brane
charge is not conserved with respect to changes in the position of the
branes. Some of these subtleties appeared originally in
\cite{Bachas:2000ik}.

The triplet of charges exists for the other forms, e.g. the six form
$F_6 = * \tilde F_4$ and are summarized in appendix B of
\cite{Aharony:2009fc}. For the flux of $F_6 =*F_4$, is is also useful
to introduce the notion of bulk charge $Q_{bulk}$ which is the total
charges carried by the bulk fields
\be Q_2^{bulk} = \int_{Y_2} {1 \over 2} G_4 \wedge G_4 \ . \ee
Then, the bulk charge can be understood as being related to the brane
and Maxwell charges via
\be Q_2^{Maxwell} = Q_2^{brane} + Q_2^{bulk} \ . \ee
To correctly quantize the supergravity solution, one should impose the
discreteness condition on the Page charges, and not on Maxwell, brane,
or bulk charges.

\subsection{Quantization on the Stiefel cone}

To illustrate the integrality of Page charges and the non-integrality
of the other charges, let us first carryout the quantization procedure
for the Stiefel cone.

First, consider the flux of $\tilde F_4$. The Stiefel geometry has
vanishing fourth Betti number, so there is no $G_4$ to consider in
M-theory, and after dimensional reduction, the IIA flux $\tilde F_4$
also vanishes.  We are not done yet, however, because we still have to
consider the Page flux (\ref{q4page}), which contains a term $B_2
\wedge F_2$, and $F_2$ is nonvanishing in the dimensional reduction of
the orbifolded Stiefel cone.  Requiring the Page flux to be integer
quantized imposes the quantization condition
\be \int B_2 = -{l \over 2k} + {1 \over 2} \ee
which we inferred independently from M-theory considerations
earlier in section \ref{sec2.1}.

Next, we consider the quantization of flux of D2 charge through
$M_2$. We are interested in determining the Maxwell charge when the
Page charge is set to $N$. One finds
\be Q_2^{Maxwell} = N - {l(l-2k) \over 2k} \label{max1}\ee
which can essentially be viewed as the sum of a contribution from $N$ M2-branes
and a contribution from the discrete torsion, along the lines of
\cite{Bergman:2009zh}. The Maxwell charge $Q_2^{Maxwell}$ has several
notable features. First, it is not necessarily integer valued. Second,
it is invariant under
\be N \rightarrow N+l, \qquad l \rightarrow l+2k \ . \ee
This is consistent with the property of Maxwell charge that it is
conserved under continuous deformations corresponding to moving one of
the 5-branes around the $S^1$ in the type IIB brane
construction. Finally, $Q_2^{Maxwell}$ can go to zero or negative for
some range of $(N, l, k)$. This suggests that the entropy of the
superconformal field theory is going to zero or negative, signaling a
phase transition. The condition that $Q_2^{Maxwell}$ is positive is
also related to the condition necessary for supersymmetry to be
unbroken as was highlighted in related contexts in
\cite{Aharony:2009fc,Hashimoto:2010bq}.

\subsection{Quantization in the Stenzel geometry}

Let us now extend our analysis of flux quantization to the case where
the Stiefel cone is deformed into the Stenzel geometry, as described
in Section \ref{sec2.3}.  To keep a general set of charges under
consideration, we will study the case where the Stenzel manifold has
been quotiented by a $Z_k$ orbifold action.

The most important feature of the geometry in the deep IR is its
singularity structure after the orbifold has been taken.  At the tip
of the deformed orbifolded cone, the geometry has the local structure
$(R^4\times S^4)/Z_k$, and in particular it has two fixed points
which we can think of as the north and south poles of the $S^4/Z_k$.
At each of the fixed points, the local geometry is $R^8/Z_k$
\cite{Martelli:2009ga}.  This geometric feature has a nice
implication.  The supersymmetry of the deformed Stenzel cone is
consistent with adding some mobile M2-branes, and we are free to move
some number of them to either of the orbifold fixed points.  Then the
theory on the M2-branes in the deep IR should simply be two copies of
the ABJM theory.

At any finite excitation energy the theory should feel the effects of
curvature and the self-dual four-form flux in the background which
break the supersymmetry from $\mathcal{N} = 6$ to $\mathcal{N}=2$.
However, for issues such as charge quantization, we should be able to
work in the extreme low energy limit and use our intuition from the
ABJM case.  In particular one might expect that it is possible to turn
on discrete torsion at each singularity, and we will see that this is
correct, although the torsion will be subject to some global
constraints.

First we will consider the type IIA reduction of this geometry. This
geometry develops a dilaton and curvature singularity near the
tip. However, because the geometry asymptotes to the Stiefel cone away
from the tip, and because quantization of Page fluxes in type IIA
description appropriately respects Gauss law/localization of charge
sources, we are able to partially constrain the discrete parameters of
the supergravity ansatz. We will then continue to consider the
geometry and fluxes near the core region from the M-theory
perspective, and identify additional constraints which further
restrict the parameters of the ansatz.

The Stenzel manifold admits the self-dual four form flux (\ref{g4}) which
can be derived from a three-form potential $C_3$
\cite{Klebanov:2010qs}
\be C_3 = m \beta + \alpha \Omega_2 \wedge d \gamma \label{3-form}
\ee
\be \beta = {a c \over \epsilon^3 \cosh^4{r \over 2}} \left[ (2 a^2 + b^2) \tilde \sigma_1 \wedge \tilde \sigma_2 \wedge \tilde \sigma_3 + {a^2 \over 2} \epsilon^{ijk} \sigma_i \wedge \sigma_j \wedge \tilde \sigma_k \right] \ , \label{beta}\ee
where $\Omega_2$ and $\gamma$ are as defined in
\cite{Martelli:2009ga}.\footnote{For the interested reader, $\gamma$
is the angular coordinate which is quotiented by the orbifold action,
and $\Omega_2$ is proportional to the geometric flux associated with
$\gamma$.}  Here we have added an exact term proportional to $\alpha$,
which does not affect the gauge invariant four-form flux.  This exact
term is present in the $AdS_4 \times V_{5,2}/Z_k$ system with discrete
torsion \cite{Martelli:2009ga} which is the UV limit of the Stenzel
solution.

In quantizing the flux of the type IIA Page flux through the four
cycle of $M_2$, we impose the condition
\be \int_{S^4} G_4 + n k \int_{\tilde S^3/Z_k} C_3 = (2 \pi) n \alpha=
-(2 \pi l_p)^3 (l-k), \qquad n=2 \label{aquant} \ee
which constrains $\alpha$.  Note that in the asymptotically conical limit,
the torsion is $Z_{2k}$-valued, and so $l$ takes integer values in the
range $0\le l \le2k-1$.

In addition to this, the flux of $G_4$ through $S^4/Z_k$ is
independently quantized to be integral.  This implies
\be \int_{S^4/Z_k} G_4 = {8 \pi^2 \over 3 \sqrt{3}k} m = (2 \pi l_p)^3
q  \ee
for integer $q$.  This constraint has no counterpart in the Stiefel
cone as neither the $S^4$ cycle nor the self-dual 4-form exist in that
case.

Now let us consider the quantization conditions that arise from
considering M-theory near the orbifold fixed points; we will show that
the expected charges at the singularities are compatible with the IIA
calculations.  At the north pole of $S^4/Z_k$, the pull-back of $G_4$
on the $R^4/Z_k$ fiber was computed in \cite{Martelli:2009ga}:
\be {1 \over (2 \pi l_p)^3} \int_{R^4/Z_k} G_4 = {q \over 2} \equiv {\tilde M \over 2} = M \ee
where $M$ and $\tilde M$ are the variables used in
\cite{Martelli:2009ga}.  This means that the integral of $C_3$
(including both the nontrivial flux and the discrete torsion
contribution) on $S^3/Z_k$ at the north pole is
\be {1 \over (2 \pi l_p)^3} \int_{\tilde S^3/Z_k} C_3 = -{l \over 2k} + {1 \over 2} -{q \over 2} \ \label{c3north}.\ee

Suppose that at the north pole we impose the condition that the system is described by charges as in the ABJ case with $l^N$ units of discrete torsion (including a shift by 1/2 a unit as discussed in \cite{Aharony:2009fc}.)  This is compatible with (\ref{c3north}) provided that
\be -{l^N \over k} + {1 \over 2}  = -{l \over 2k} + {1 \over 2} -{q \over 2} \ee
or equivalently
\be l^N = {l + kq \over 2}. \label{lN}\ee

At the south pole, the computation is very similar, except that the
flux quantum $q$ appears with a minus sign:
\be l^S = {l - kq \over 2} \label{lS}\ee
The difference in the pull-back of $C_3$ between the north and the
south pole is just the total flux $q$, while the discrete torsion
contribution must be the same at the north and south poles because the
torsion has no associated flux.\footnote{In the coordinates of
\cite{Bergman:2001qi,Klebanov:2010qs}, the $U(1)_b$ quotient as
defined in \cite{Martelli:2009ga} is imposed on the angular coordinate
$\phi_2$.  With this choice of $U(1)$ action, the poles of the
$S^4/Z_k$ are located at $(\tau=0,\alpha = \pi/2,\psi = 0,\pi)$.  In
the vicinity of the poles, one can check that the one-forms
$\tilde{\sigma_i}$ differ by a sign,
$\tilde{\sigma}_i(N)=-\tilde{\sigma}_i(S)$, so the three-form $\beta$
in (\ref{beta}) also changes by a sign from the north pole to the
south pole.}

How should we interpret the formulas (\ref{lN}) and (\ref{lS})?  The
first thing to note is that $l^N$ and $l^S$ are equal mod $k$, so if
they had described decoupled systems we would have said that they were
equivalent up to a large gauge transformation.  However, they are not
decoupled, and there is no large gauge transformation that sets them
equal to each other.  Instead, the picture that has emerged is that
$l^N$ and $l^S$ locally appear to describe the same torsion, but
globally there is a topologically nontrivial twist relating them, and
the winding number of the twist is just the number of units of $G_4$
flux in $S^4/Z_k$.

The second thing to note is that in the local ABJ models at the north
and south poles, $l^N$ and $l^S$ should themselves be integers, or in
other words $l-kq$ must be even.  This means that for a given $q$ and
$k$, $l$ must take either only even or only odd values.  In the
undeformed theory, $l$ described a $Z_{2k}$-valued discrete torsion,
but we see that our local considerations at the tip of the Stenzel
geometry remove half of the possible values of $l$, and the discrete
torsion in the deformed case is $Z_k$-valued.  This phenomenon is
reminiscent of the deformed conifold; the ``singular'' conifold admits
a $Z_2$-valued discrete torsion which is not present in the deformed
conifold \cite{Vafa:1994rv}.

We can now examine the quantization of the six form flux though $M_2$
in IIA or the 7-form flux through $V_{5,2}$ in M-theory which measures
the charge of D2/M2 branes in this background.

One way to approach this issue is to first examine the brane charges
present in this setup. Before adding any explicit 2-branes, there are
2-brane charges arising from the discrete torsion at the $Z_k$ fixed
points at north and south poles \cite{Bergman:2009zh}. These should
have the same form as what was computed in \cite{Aharony:2009fc}, so
we find
\be Q_2^{torsion} = \left(-{l^N \over k} + {1 \over 2}\right)+\left(-{l^S \over k} + {1 \over 2}\right) = -{l(l-2k) \over 4k}- {k q^2 \over 4} \ . \ee
If, in addition, we were to introduce $N$ 2-branes which can be at any
point in the Stenzel geometry, there will be an additional
contribution of $N$ to the brane charge
\be Q_2^{brane} = N-{l(l-2k) \over 4k}- {k q^2 \over 4} \ . \ee
Since Maxwell charge is the sum of brane charge, and since the bulk
charge is given by
\be Q_2^{bulk} ={1 \over (2 \pi l_p)^6} \int_{{\cal M}_8} {1 \over 2} G\wedge G =  {2^{11} m^2 \mbox{vol}(V_{5,2}) \over (2 \pi l_p)^6 3^6} = {k q^2 \over 4}\ee
we infer that
\be Q_2^{Maxwell}=N-{l(l-2k) \over 4k} \ .  \label{max2}\ee
It also follows that the Page charge $Q_2^{Page}=N$.

This result is gratifying for several reasons. First, this result
reflects the accounting of all identifiable charge sources in an
otherwise consistent and smooth M-theory background aside from the
orbifold fixed point. The final answer is the same as what we inferred
for the undeformed Stiefel cone (\ref{max1}). It then follows that
the gauge invariant Maxwell charge is invariant under the shifts
\be N \rightarrow N+l, \qquad l \rightarrow l + 2k\ee
which arises naturally from several perspectives mentioned earlier.

The only additional constraint imposed by the Stenzel deformation is
the restriction on the parity of $l$ so that $l$ is congruent to $kq$
mod 2. This is far milder than what was found in
\cite{Martelli:2009ga}. 

\section{Discussion}

In this article, we reviewed the quantization of fluxes in warped
Stiefel cone and its Stenzel deformation which is conjectured to be
the holographic dual of ${\cal N}=2$ Chern-Simons matter theory in 2+1
dimensions.  We described the subtle difference between several
different yet related notions of charges, and recovered a structure
compatible with the pattern of Hanany-Witten brane creation effects
and duality cascades.

There are a number of interesting features which one can infer from
the structure of the gravity solution. $Q_2^{brane}$ is a measure of
the number of degrees of freedom in the deep infrared of this
system. When $Q_2^{brane}$ is zero or negative, we expect the system
to break supersymmetry and flow to a different universality class of
vacuum as was the case for many related system
\cite{Aharony:2009fc,Hashimoto:2010bq}. It would be very interesting
to better understand the nature of the effective low energy physics
when the system is in this new phase. This question can be addressed in
the simple context of $k=1$ where there are no $Z_k$ orbifold fixed
points, and by taking $q$ to be even, we can even set $l=0$ and
disregard the contribution from the discrete torsion.

\begin{figure}
\centerline{\begin{tabular}{ccc}
\includegraphics[width=1.8in]{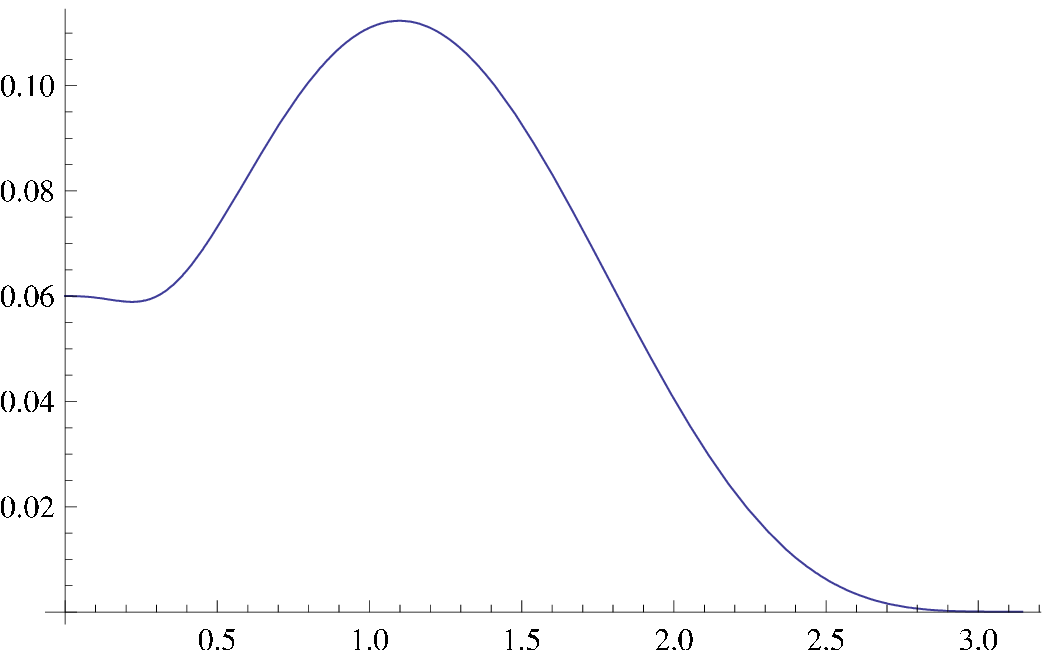} & 
\includegraphics[width=1.8in]{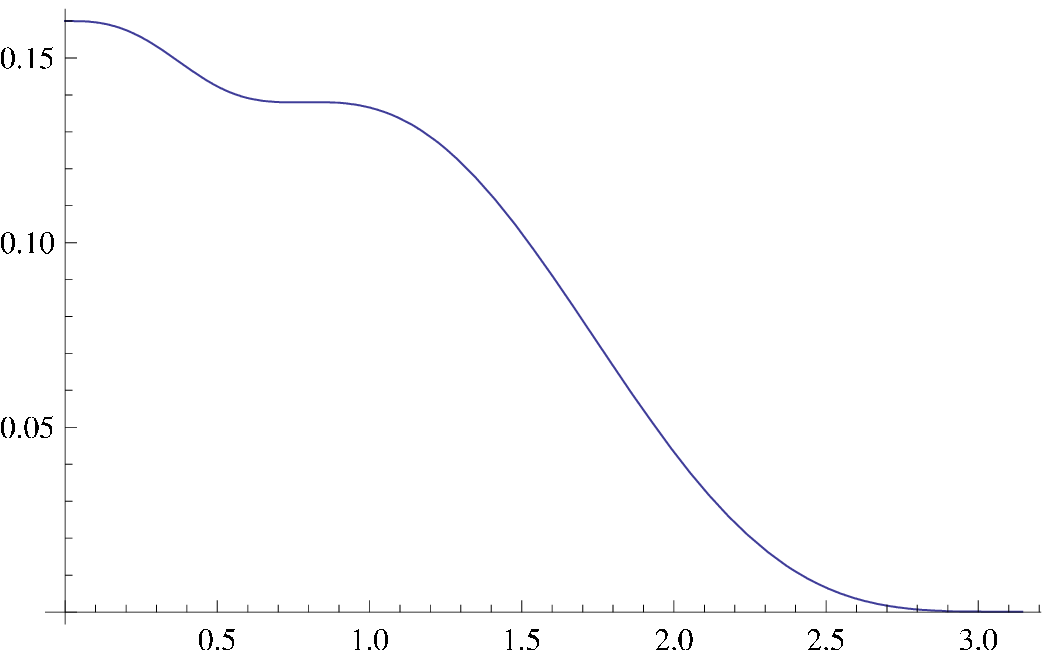} & 
\includegraphics[width=1.8in]{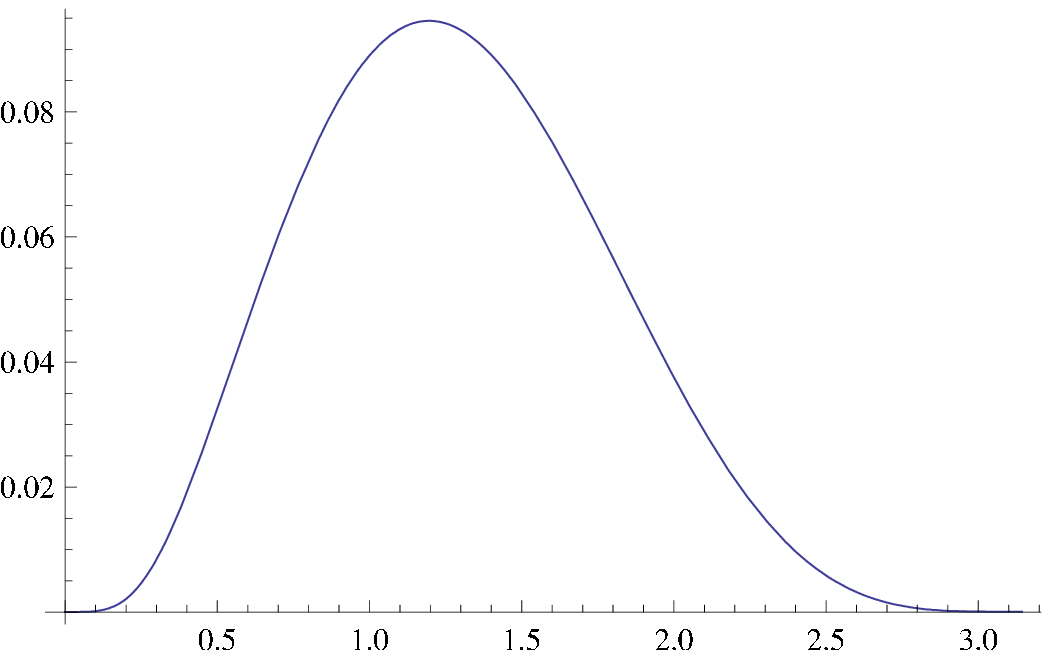} \\
(a) & (b) & (c)
\end{tabular}}
\caption{Potential $V(\psi)$ for $p$ anti D3-brane blowing up to an
NS5-brane wrapping an $S^2$ of fixed latitude in $\psi$ in $S^3$ at
the tip of the Klebanov-Strassler solution. (a), (b), and (c)
corresponds to $p/M=0.03$, $p/M=0.08$, and $p/M=-0.03$, respectively.
These figures originally appeared in figure 2 of \cite{Kachru:2002gs}.
\label{figa}}
\end{figure}

One way to probe the fate of pushing the system which is slightly
perturbed into this new phase is to start with a background with $q$
large but $Q_2^{brane}=0$ (which can easily be arranged for $q$ even
and $k=1$). Consider now adding $p\ll q$ anti M2-brane as a
probe. This setup is very similar to adding anti D3-brane in warped
deformed conifolds \cite{Kachru:2002gs} which has received a lot of
attention (and controversy) as a possible prototype as a gravity dual
of a metastable vacua
\cite{DeWolfe:2008zy,Bena:2009xk,Bena:2010gs}. For the Stenzel
manifold, the effective action of the brane probe undergoing a KPV-like
transition \cite{Kachru:2002gs} works essentially in the same way as
is illustrated in figure \ref{figa}. However, from the point of view
of the bound $Q_2^{brane}>0$, one expects the stable supersymmetric
minima not to exist when $p$ anti M2-branes are introduced. 

Tentatively, we interpret these facts as follows.  The computation of
the potential $V(\psi)$ neglected the backreaction of the anti-branes,
and when the number of anti-branes is parametrically small ($p \ll q$)
this probe approximation is valid.  In particular, the existence of
the non-BPS local minimum in \ref{figa}.(a) is a robust prediction in
this limit.  However, when the state in the metastable false vacuum
illustrated in figure \ref{figa}.(a) tunnels to the putative ``true''
vacuum, the amount of charge carried by the probe grows to $q-p$ which
is not parametrically small compared to $q$.  The backreaction due to
this charge can be significant, and so the computation of the
tunneling potential is (at least) not obviously self-consistent.  It
is tempting to speculate that the supersymmetric vacuum might actually
be spurious and that the non-BPS local minimum is the global minimum
which characterizes the dynamics in the $Q_2^{brane}<0$ phase of these
theories up to corrections suppressed by $p/q$.

Similar considerations apply to the BPS domain wall one constructs for $p
< 0$ for which the KPV potential has the form illustrated in figure
\ref{figa}.(c). This domain wall can also be viewed as arising from
wrapping an M5-brane on $S^4$ at the tip of the Stenzel manifold. A
5-brane wrapped on a 4-cycle is effectively a string, and in 2+1
dimensions, a string forms a domain wall. It would be very interesting
to understand the nature of vacua separated by these domain
walls. Since M5-brane wrapped on $S^4$ with $q$ units of flux must
have $q$ additional M2-branes ending on it to cancel the anomaly, some
quantum numbers of the vacuum must shift to reflect this. Nonetheless,
one expects the Maxwell charges $Q_2^{Maxwell}$ and $Q_4^{Maxwell}$ to
be invariant as one crosses the domain wall, as these charges are
conserved. Making complete sense of these expectations requires taking
the full back reaction of the M5-brane and the $q$ anomalous M2-branes
into account. Unfortunately, $q$ M2-branes can not be treated reliably
as a probe, making systematic analysis of these issues a challenge.

Let us also mention that similar issues of stable/metastable non-BPS
vacua, domain wall, and low energy effective field theories can be
discussed in the closely related $B_8$ system building on the analysis
of \cite{Hashimoto:2010bq} and \cite{Gukov:2001hf}. Quantization of
charges and the enumeration of brane, Maxwell, and Page charges for
this system was carried out in \cite{Hashimoto:2010bq}. Here, however,
we encounter one additional puzzle. It was argued in
\cite{Gukov:2001hf} that the 4-form flux through $S^4$ at the tip of
the $B_8$ cone is half integral as a result of the shift originally
due to Witten \cite{Witten:1996md}. This would appear to require half
integer units of M2-branes to end on the domain wall made by wrapping
the M5 on the $S^4$. Of course, the number of M2's ending on an M5 is
constrained to be an integer.  Perhaps this is indicating that odd
number of M5-branes are forbidden from wrapping the
$S^4$. Alternatively, this paradox is another manifestation of not
systematically taking the back reaction of the domain wall into
account.

Finally, let us emphasize that for the time being, the concrete field
theory interpretation of the Stenzel deformation and the quantum
number $q$ is not known.  The gravity dual suggests that the parameter
$q$ is important for both the IR and the UV physics.  At large radius,
$q$ is related to the total number of units of M2-brane charge
generated by the cascade, which in turn affects the UV gauge symmetry.
Near the tip of the Stenzel geometry, the $G_4$ flux is nonvanishing
so $q$ should also appear in the data of the IR field theory.  Of
course, $q$ can only be nonzero when the geometry is deformed.
Martelli and Sparks conjectured that this deformation was related to
turning on a particular mass term on the field theory side.  One can
indeed see that the null geodesic can travel from boundary at infinity
to the core in finite field theory time, and so the spectrum of
glueball-like states will exhibit a discrete structre whose scale is
set by the deformation.  If this conjecture is correct, it would
suggest that the field theory confines because of a mass deformation
(reminiscent of the $\mathcal{N}=1^*$ theory in
$d=4$\cite{Polchinski:2000uf} and the mass deformed ABJM theory
\cite{Lin:2004nb,Gomis:2008vc,Kim:2010mr,Cheon:2011gv}) rather than as
a dynamical effect, as is the case in the Klebanov-Strassler system
\cite{Klebanov:2000hb}. It should be very interesting to understand
this theory better.

\section*{Acknowledgements}

We would like to thank  Ofer Aharony and Shinji Hirano for collaboration on
related issues and for discussions at the early stage of this work.
We also thank 
Igor Klebanov,
Don Marolf,
Dario Martelli, and
James Sparks
for useful comments and discussions. The work of AH is
supported in part by the DOE grant DE-FG02-95ER40896 and PO is supported in part by DOE grant
DE-FG02-91ER40681.

\appendix

\section{Charge quantization and duality transformations}

One of the subtle features arising in quantizing the supergravity
background in this article is the fact that some fluxes lifts or
dualizes to a quantity encoded not by the period of a field strength,
but by a quantity like the discrete torsion which is the period of a
potential field. In this appendix, we illustrate several examples, in
a simpler context, giving rise to similar subtleties.

\subsection{Charge quantization for duals of $TN \times S^1$}

The KK-monopole, also known as the Taub-NUT space, is a well known
Ricci-flat gravitational background.  The metric for $TN \times S^1$
has a simple form
\be \left(  1 + {R_1 \over 2r}\right) (dr^2 + r^2 (d \theta^2 + \sin^2 \theta d \phi^2)) + {R_1^2 \over \left(1 + {R_1 \over 2r}\right)} \left(d \psi + {1 \over 2} \cos \theta d \phi\right)^2 + R_2^2 d \eta^2 \label{KK5S1}\ . \ee
We take $\phi$, $\psi$, and $\eta$ to have period $2 \pi$, and $0 \le
\theta \le \pi$. $R_1$ and $R_2$ are the radius of $S^1$ parametrized
by $\psi$ and $\eta$, respectively.

Being Ricci flat, this metric can easily be embedded in
M-theory. Reducing to IIA along $\eta$ will give rise to a KK5-brane
in type IIA string theory. Reducing to IIA along $\psi$ will
give rise to a D6 brane extended along $\eta$.

Consider a general linear transformation on the coordinates $\psi$ and
$\eta$
\be \eta = a \eta' + b \psi', \qquad \psi = c \eta' + d \psi' \label{generallinear}\ . \ee
This will modify the last two terms  of (\ref{KK5S1}) to
\be {R_1^2 R_2^2 \over (c^2 R_1^2 + a^2 R_2^2 V)} \left((ad-bc) d \psi' -  {a \over 2} \cos\theta d \phi\right)^2  
 + \left(a^2 R_2^2 + {c^2 R_1^2 \over V}\right) \left( d\eta' + A_1
\right)^2 \ee
with
\be A_1  
 =  
\left({c R_1^2 \over a (c^2 R_1^2+a^2 R_2^2V)} +{1 \over (ad - bc)}{b \over a}\right) \left((ad-bc)d \psi' + {a \over 2} \cos \theta d\phi\right)-
 {1 \over (ad-bc)}{b \over 2} \cos \theta d\phi \label{generala1} \ . 
\ee

At the level of classical supergravity, this is a solution generating
transformation, but not all of the solutions obtained in this fashion
are sensible backgrounds of string theory.  Rather, there is a certain
discrete subset of these solutions which is consistent with charge
quantization.

\subsection{Twisted $Z_p$ orbifold\label{secA2}}

One example of such a discrete subset is to take
\be \left(\begin{array}{cc} a & b \\ c & d \end{array}\right) = \left(\begin{array}{cc} 1 & {q/p} \\ 0 & 1/p \end{array}\right) \ee
and impose the periodicity $\eta'=\eta' + 2 \pi$ and $\psi' = \psi' +
2 \pi$. This can be viewed as a twisted $Z_p$ orbifold of
$TN\times S^1$ as outlined in (3.4) of \cite{Witten:1997kz}. When
reducing to IIA, one finds a RR 1-form potential
\be A_1 = \left({q \over p} d \psi' + {q \over 2} \cos \theta d \phi\right) - {q \over 2} \cos \theta d \phi = {q \over p} d \psi' \label{a1}\ . \ee
Upon further T-dualizing this background along the $\psi'$ direction,
we obtain a supergravity solution for a $(p,q)$ 5-brane\footnote{The
$p$ and $q$ are switched from what is in \cite{Witten:1997kz} so that
$p$ counts the number of NS5-brane and $q$ counts the number of
D5-brane in the dual IIB description.} smeared along the $\psi'$
direction. This can easily be seen from the RR 2-form potential
\be A_2 = - {q \over 2} \cos \theta d \phi \wedge d \tilde \psi' \ee
and the NSNS 2-form potential
\be B_2 = - {p \over 2} \cos \theta d \phi \wedge d \tilde \psi' \ee
which one finds from the duality transformation. The 3-form field
strength is closed and naturally encodes the flux through $S^2 \times
S^1$ parametrized by $\theta$, $\phi$, and $\tilde \psi'$ for
arbitrary $r$.

The issue stems from attempting to understand the $q$ units of D6
charge from the IIA description {\it prior} to taking the T-duality
along the $\psi'$ direction. One expects the D6 charge to be encoded
by the flux of the RR 2 form field strength
\be F_2 = d A_1 \ee
but this {\it vanishes} for the background (\ref{a1}) under
consideration. For this example, the hint for where the quantum number
of the D5 charge is encoded in the IIA description is staring at our
face in equation (\ref{a1}).  It is the fractionally valued vector
potential arising as a result of the non-trivial twist, $q$, in the
$Z_p$. This may be thought of as the simplest example illustrating the
point that charge is sometimes encoded in the period of a potential,
i.e.  a Wilson line, rather than the field strength.

\subsection{$SL(2,Z)$ dual of $TN \times S^1$\label{secA3}}

Let us now consider a different example, where we take the $SL(2,Z)$
subset of the general linear transformation (\ref{generallinear}). In
this case, we have $ad-bc=1$, simplifying (\ref{generala1}) to
\be A_1  
 =  
\left({c R_1^2 \over a (c^2 R_1^2+a^2 R_2^2V)} +{b \over a}\right) \left(d \psi' + {a \over 2} \cos \theta d\phi\right)-
{b \over 2} \cos \theta d\phi \label{sl2za1} \ . 
\ee
Once again, T-dualizing on $\psi'$ will give rise to a RR 2-form
\be A_2 = - {b \over 2} \cos \theta d \phi \wedge d \tilde \psi' \ . \ee
In fact, if we take $a=p$ and $b=q$, the IIB 5-brane charges are
identical to the example in the previous section although the
background differs in the asymptotic value of the axiodilaton.

The puzzle, once again, is the status of the D6 charge in the IIA
frame. This time, the RR 2-form field strength does not vanish, so one
might try to define a D6-brane charge by integrating $F_2$ over a
suitable 2-cycle.  However, no such nonsingular 2-cycle exists.  For
example, integrating on the natural $S^2$ parametrized by $\theta,
\phi$ would give a charge that depends on the radius $r$.  This
apparent failure of the Gauss law can be traced to this $S^2$ not
actually being a well-defined 2-cycle.

Notice that in the ordinary IIA reduction of (\ref{KK5S1}) on the
circle parametrized by $\psi$, the procedure of integrating $F_2$ on
the $S^2$ at fixed radius is the correct one for counting the number
of D6 branes.  On the other hand, for the IIA reduction on $\eta$ (or
in M-theory) the integrality of the D6 brane charge follows from
demanding that the KK5 metric does not have a singularity.  In a
generic duality frame, such as the one given by reduction on $\psi'$,
neither condition is correct.

Instead, one might try to define a modified flux quantization
condition that mixes the flux and geometry in such a way as to obtain
a conserved charge.  This can be done by considering the combination
\be  Q_{D6}={1 \over 2 \pi} \left(\int_{S^2} F_2 + {a \over ad - bc} \int_{S^1} A_1\right)  = {b \over ad-bc}\ee
where $S^2$ is the 2-sphere parametrized by $(\theta,\phi)$, and
$S^1$ is the circle parametrized by $\psi'$. Note that this reduces
to the same prescription as in the previous subsection if $F_2$
happens to vanish. In fact, one could have also considered applying
the prescription of reading off the period of $A_1$ precisely at the
radius $r=0$ where $F_2$ would have vanished.

Formally, this procedure is equivalent to considering
\be  Q_6 = {1 \over 2 \pi} \int_{S^2} \left(F_2 + f \wedge A_1\right) \ee
discussed briefly in appendix A of \cite{Aharony:2009fc} based on the
language of
\cite{Kachru:2002sk,Shelton:2005cf,Ellwood:2006ya,Villadoro:2007tb}. Although
the procedure of computing the period of this ``modified'' flux will
give the correct answer, it is somewhat unsettling that the procedure
is not generally covariant. The aim of this article is to identify the
origin of this charge from discrete data of the potential.

\subsection{Quantization of gravity duals of field theories in 2+1 dimensions}

Let us now examine the quantization of fluxes in gravity duals of
various 2+1 dimensional field theories and compare their features with
the example of the previous section.

We will begin by reviewing the case of ABJM and ABJ for which the
gravity theory is M-theory on $AdS_4 \times S^7 / Z_k$
\cite{Aharony:2008ug,Aharony:2008gk}. The transverse $S^7 / Z_k$ has a
torsion 3-cycle on which one can wrap $l$ M5-branes in the range $0
\le l <k$. These M5-branes will not source M-theory 4-form
flux. Instead, they give rise to discrete torsion parametrized by a
flat $C_3$ in the background, supported on the $S^3/Z_k$. This is
quite similar to the case of the one form potential (\ref{a1}) in the
case of twisted $Z_p$ orbifold of $TN \times S^1$ we described in
section \ref{secA2} and \ref{secA3}. There are two refinements to this
story.

One is that the quantization condition for the discrete torsion has an
anomalous shift due to the Freed-Witten anomaly, and reads
\be k \int_{S^3/Z_k} C_3  = l - {k \over 2}  \ . \label{a14} \ee
The presence of Freed-Witten anomaly was inferred in the IIA
description of this background in \cite{Aharony:2009fc}. At the
moment, it is not clear how one understands this shift strictly in the
M-theory perspective, but since it is required in the IIA reduction,
we will adopt it in the M-theory lift as well. One can simply view
this as an overall shift in the charge lattice. As we will see below,
this shift turns out to be consistent with a rather non-trivial
consistency test.

The second refinement concerns the relation between the M2 charge and
the radius of anti de Sitter geometry. In the absence of discrete
torsion, the radius of anti de Sitter space is directly proportional
to the number of M2 branes giving rise to the near horizon $AdS_4\times
S^7$ geometry. In the presence of discrete torsion, however, the
relation receives a correction. This issue was investigated originally
in \cite{Bergman:2009zh} which left out the contribution from the
Freed-Witten anomaly. Taking the Freed-Witten anomaly into account
\cite{Aharony:2009fc}, one finds that
\be R^4 = (2^5 \pi l_p^4){ Q_2 \over k} \ee
where
\be Q_2 = N - {l (l-k) \over 2 k} + Q_{curv} \ee
with 
\be Q_{curv} = -{1 \over 24} \left( k - {1 \over k}\right) \ee
is the contribution from the $C_3 \wedge R^4$ correction to the
M-theory action.

In the IIA reduction along the Hopf fiber of $S^3/Z_k$, the $Q_2$ can
also be written
\be Q_2 = \left(N + {k \over 8}\right) +  \left(l - {k \over 2} \right) b + {1 \over 2} k b^2 + Q_{curv}, \qquad b = -{l\over k} + {1 \over 2} \label{qmaxb} \ee
where $b$ is the pull-back of $B_2$ on $S^2$ level surface of
$R^4/Z_k$ reduced on $S^1$. This expression makes the interpretation
of $Q_2$ as including the contribution from the $B$-field in the
Wess-Zumino term for $k$ D6-brane and $l-k/2$ D4-branes, including the
Freed-Witten shift, manifest. There is also a fractional shift in the
D2 charge by $k/8$ and from $Q_{curv}$. The $k/8$ shift can be viewed
as following from the Freed-Witten anomaly.

Recently, in a very impressive paper \cite{Drukker:2010nc}, the planar
free energy of the ABJM/ABJ field theory was computed in the strong
coupling limit on the field theory side using localization and matrix
model techniques. They found that up to numerical constants (which
they also compute), the free energy is proportional to
$Q_2^{3/2}$. This is a rather non-trivial test for the consistency of
the details of the shifts in curvature due to discrete torsion,
including the detailed form of the effects of Freed-Witten
anomaly. They in fact confirm specifically the presence of a shift in
the D2 charge by $k/8-k/24=k/12$ units.\footnote{The analysis of
\cite{Drukker:2010nc} was in the planar limit $k \gg 1$ where $N/k$
and $l/k$ were kept fixed. It is possible that $Q_2$ has corrections
suppressed by $1/k$ that has not yet been accounted for in the gravity
side\label{footnote2}.}

Let us also comment in passing that the positivity of the anti
de-Sitter radius (excluding the contribution from the curvature) is
equivalent to the condition for preserving supersymmetry implied by
generalized $s$-rule \cite{Aharony:2009fc,Hashimoto:2010bq}.

Finally, let us discuss the generalization of (A.14) when the
background supports non-trivial $G_4$, as is the case for the Stenzel
geometry.  In this case, the pull-back of $C_3$ can vary in such a way
that can be cancelled by the pull-back of $G_4$. By subtracting this
contribution, one arrives at (\ref{aquant}) which can be interpreted
as the M-theory lift of the Page charge (\ref{q4page}).

\bibliography{stenzel}\bibliographystyle{utphys}

\end{document}